\documentstyle[preprint,aps,epsfig]{revtex}
\tightenlines

\begin{document}
\date{\today}
\title{Criticality in confined ionic fluids}
\author{J. E. Flores-Mena$^{1,2}$,  
Marcia  C. Barbosa$^{2,3}$ and Yan Levin$^{2}\footnote{Corresponding author: 
E-mail: levin@if.ufrgs.br}$\\
$^1$Facultad de Ciencias de la Electr\'onica, \\
Universidad Aut\'onoma
de Puebla, A.P. J-48, Puebla 72570, Pue. Mexico\\
$^2$Instituto de Fisica, Universidade Federal do Rio Grande do Sul, \\
Caixa Postal 15051, 91501-970 Porto Alegre (RS), Brazil\\
$^3$Center of Polymer Studies, Boston University, Boston, 02215, MA, USA}
\maketitle
\begin{abstract}

A theory of a confined two dimensional electrolyte is presented.
The positive and negative ions, interacting by a $1/r$ potential,
are constrained to move on an interface separating two 
solvents with dielectric constants $\epsilon_1$ and $\epsilon_2$.
It is shown that the  Debye-H\"uckel type of theory predicts that the
this $2d$ Coulomb fluid should undergo a phase separation
into a coexisting liquid (high density) and gas (low density) phases.
We argue, however, that the formation of polymer-like chains of alternating
positive and negative ions can prevent this phase transition from
taking place.

\end{abstract}
\centerline{{\bf PACS numbers:}64.70.-p, 05.70.Fh, 64.60.-i}
\section{Introduction}

Over the last decade the need to understand coulombic
criticality has provided a new impetus to the study of
electrolyte solutions.  The current wave of exploration can
be traced to the pioneering experiments of K. S. Pitzer {\it et. al.}, 
who have first reported a surprising finding that Coulomb
interactions might belong the mean-field universality 
class~\cite{sin90}.  This suggestion has not gone unchallenged and, in
fact, later experiments are consistent with a
crossover from  mean-field to Ising universality class
very close to the critical point~\cite{wei92}. 
The crossover, if indeed it exists, 
is  much closer to the critical point
than for any other known fluid.

From the theoretical point of view it is very hard to justify
anything but Ising criticality~\cite{fis93,lev96}.  The goal for the
theorists must then lie in a seemingly simpler task of finding
why the crossover region for electrolytes is so narrow.
Unfortunately even this programme has failed to produce
any satisfactory explanation~\cite{fis96}.  Most calculations suggest that
the critical region for electrolytes should be comparable
to that of a Lennard-Jones fluid.  To further confound
the mystery, Monte Carlo simulations are once again pointing
in the direction of mean-field criticality~\cite{val98}.

A seemingly unrelated problem concerns the disappearance of
the anticipated liquid-gas transition in a system of dipolar hard
spheres (DHS). Since the DHS is the simplest realization of a 
polar fluid, for a long time it has been believed
that it must exhibit a liquid-gas phase separation.  
It came, therefore, as quite
a surprise when the Monte Carlo simulations failed to located this 
singularity~\cite{wei93}.  Instead what they found  was that as
the temperature was lowered, the dipolar particles aligned,
forming polymer-like chains.  Since these chains interact
weakly~\cite{sea96}, it has been argued that the liquid-gas 
transition must be driven
entirely by the free unassociated dipoles~\cite{lev99}.  In fact a critical
density of free dipoles is needed for the phase transition
to occur.  At low temperatures, 
where most of the theories localize
the transition, formation of dipolar chains 
depletes the density of free dipoles bellow the critical threshold
necessary for the transition to take place~\cite{lev99}.

In this paper we shall study a third model which we hope
might span the gap between the three dimensional Coulomb gas
and the dipolar hard spheres, and thus shed some additional
light on the criticality in these interesting and important 
systems.  
Our new model  consists of a neutral
electrolyte confined to a two dimensional plane~\cite{vel97}.  This can be
visualized as oppositely charged surfactant molecules
adsorbed to a water-oil interface. An example of such a system is
cetyltrimethyl amonium-hydroxy naphthalene 
carboxylate $(CTAHNC)$, which is composed of two surface
active parts, $CTA^+$ and $HNC^-$~\cite{sal96}.

We shall argue that unlike the $3d$ electrolyte~\cite{val91}, the confined
$2d$ plasma might not phase separate.  Instead as the temperature
is lowered, chains composed of alternating positive and
negative ions will begin to form $(...+-+-...)$~\cite{lev96}.  Just as dipolar
chains, these new clusters interact weakly between 
themselves. However, they diminish the concentration 
of free ions bellow the critical value necessary for the transition
to take place.

\section{The Model}

Our system consists of an ionic fluid of $N_+$   
positive and $N_-$ negative ions
confined to a plane of area $A$ located at $z=0$, separating two
different solvents occupying the half-spaces at  $z>0$ and $z<0$. 
We shall restrict our attention to the neutral electrolytes,
for which $N_+=N_-=N/2$. The  
solvents are treated  as uniform mediums with dielectric
constants $\epsilon_{2}$ and $\epsilon_{1}$ for $z>0$ and $z<0$, 
respectively. The  ions of both species are modeled as hard
spheres of diameter $a$ and charge $\pm q$ located at the center.
It will be convenient to define 
the dimensionless particle density as
$\rho_{\pm}^{*}=\rho_{\pm} a^{2}$, where $\rho_i=N_i/A$. The total
density of ions is $\rho=\rho_++\rho_-$.

All the relevant thermodynamic information is contained in the
free energy density, $f=F/A$.  Unfortunately due to the
complexity of interactions, no exact expression for
$f$ can be found.  We shall, therefore, attempt to
construct the approximate free energy using the most
relevant contributions.  
These can be divided into entropic and electrostatic,
\begin{equation}
f(T,\rho_+,\rho_-)=  f_{ent}^{(1)}(T,\rho_+,\rho_-)+ 
f_{el}(T,\rho_+,\rho_-)\; .
\label{fDH}
\end{equation}
The entropic (mixing) free energy is given by
\begin{eqnarray}
\label{1}
\beta f_{ent}^{(1)}(T,\rho_+,\rho_-)&=&  \left[  \rho_{+} \ln( \rho_{+} 
\Lambda^{2} ) - \rho_{+} + \rho_{-} \ln( \rho_{-} 
\Lambda^{2} ) - \rho_{-}\right]
\end{eqnarray}
\noindent where $\Lambda= \sqrt{2 \pi \hbar^{2}/ m k_{B}T}$ 
is the thermal de Broglie wavelength of the ions.
The second term in Eq.~(\ref{fDH}) is due to  electrostatic interactions.
It is important to note that the electrostatic free energy is purely
correlational, since the mean-field contribution is zero.
To calculate $f_{el}$, let us fix one ion at the origin.
Adopting the cylindrical coordinate system $(\varrho, \varphi, z)$, 
the central ion is located at $\varrho=0, \;z=0$. 
Due to the electrostatic interactions the other
particles will arrange themselves {\it within the plane}
in accordance with the Boltzmann distribution.
Since no charge is present in the regions  $z>0$ and $z<0$, 
the electrostatic potential there must satisfy the Laplace
equation.  Appealing to the azimuthal symmetry and taking
into account the fact that the potential should vanish  
at infinity  we find~\cite{lev299},   
\begin{equation}
\phi_{1}(\varrho,z) = \int_{0}^{\infty} A_{1}(k) 
J_{0}(k\varrho){\bf{\mbox{\large{\em e}}}}^{k z}\; dk\;\;\;\; 
for \;\;\; z<0
\label{2}
\end{equation}
\noindent and
\begin{equation}
\phi_{2}(\varrho,z) = \int_{0}^{\infty} A_{2}(k) 
J_{0}(k\varrho){\bf{\mbox{\large{\em e}}}}^{-k z}\; dk\;\;\;\; 
for \;\;\; z>0 \;\;\;,
\label{3}
\end{equation}
\noindent where $J_{0}(x)$ is the Bessel function of order zero.

The functions $A_1(k)$ and $A_2(k)$ are determined by the boundary conditions:
continuity of electrostatic  potential, 
\begin{equation}
\phi_{2}(\varrho,0)=\phi_{1}(\varrho,0) \;\;,
\label{4}
\end{equation}
\noindent and discontinuity of 
displacement field across the $z=0$ plane,
\begin{equation}
\left( \epsilon_2{\bf E}_{2}(\varrho, z) - \epsilon_1{\bf E}_{1}(\varrho,z)  
\right) \cdot \hat n = 4\pi \sigma_{eff}(\varrho) \;\;,
\label{5}
\end{equation}
\noindent  where $\sigma_{eff}(\varrho)$ is the surface
charge density and $\hat n$ is a unit vector
normal to the interface, pointing from region 1 to 2. The continuity of 
electrostatic
potential results in  $A_{1}(k)=A_{2}(k)=A(k)$, while
Eq.~(\ref{5}) requires that
\begin{equation}
2\int_{0}^{\infty}k  A(k) J_{0}(k \varrho) dk =\frac{4\pi
\sigma_{eff}(\varrho)}{D} \;\;\;, 
\label{6} 
\end{equation}
\noindent where $D=(\epsilon_1+\epsilon_2)/2$.
The surface charge  density~\cite{lev299} is given by
\begin{equation}
\sigma_{eff}(\varrho)=\sigma_{s}(\varrho)+\frac{q\delta(\varrho)}{2\pi
\varrho}\;. \label{7}
\end{equation}
The term $\sigma_{s}$ is the charge density 
of the "ionic cloud" around the central ion,
\begin{equation}
\sigma_{s}(\varrho)=q\rho_+ e^{-\beta q \phi}- q\rho_- e^{+\beta q \phi}.
\label{8}
\end{equation}
\noindent Eqs.~(\ref{6})  and (\ref{7})  are only 
valid  in the limit $a \rightarrow 0$.  
Unfortunately it is highly
nontrivial to take into account the boundary condition
associate with the finite ionic size.  To circumvent this difficulty, 
we shall first solve all the equations in the point particle limit.  
Then to account
for the finite particle size, we shall replace the bare charge of the
central ion $q$, in Eq.~(\ref{7}), by an effective charge $Q$,
$q \rightarrow Q$. The effective charge $Q$ will be determined
by the condition of an overall charge neutrality,
\begin{equation}
\label{8a}
 2 \pi \int_a^{\infty} \sigma_s(\varrho) \varrho d \varrho = -q. 
\end{equation} 
 
For the $3d$ electrolyte this procedure leads to 
an electrostatic potential identical with the one found from 
the exact solution of the Debye-H\"uckel equations with the 
appropriate hard core  boundary
conditions~\cite{deb23}.
In the present geometry the procedure outlined above 
will only be an approximation, although we believe a rather good
one.  

In the spirit of Debye-H\"uckel theory~\cite{deb23} we shall now linearize the
Boltzmann factor in Eq.~(\ref{8}). The surface charge density then 
reduces to
\begin{equation}
\sigma_{s}(\varrho)= -\frac{D\kappa_{s}\phi(\varrho)}{2\pi} \;\;\;,
\label{9}
\end{equation}
\noindent where  $\kappa_{s}=2 \pi \rho^*/ T^* a$ 
is the inverse Gouy-Chapman length and  $T^*=k_BTDa/q^2$ is
the reduced temperature.
Eq.~(\ref{6}) can now be solved yielding the expression for
electrostatic potential,
\begin{equation}
\phi (\varrho,z)=\frac{Q}{D}\int_{0}^{\infty}\frac{k}{k+\kappa_{s}}J_{0}(k\varrho)
{\bf{\mbox{\large{\em e}}}}^{-k\vert z \vert} dk.
\label{11}
\end{equation}
\noindent For $z=0$, the in plane potential agrees 
with the one obtained by Velazquez and Blum~\cite{vel97} and can  
be conveniently rewritten as
\begin{equation}
\label{>}
\phi_{>}( \varrho )= \frac{  Q \tau_{0}(\kappa_{s}\varrho) }
 { D \varrho }, 
\label{12}
\end{equation}
\noindent The subscript "$>$" is included to stress that for rigid
particles this form will be appropriate 
only outside the hard core exclusion region, 
$\varrho>a$. The function $\tau_{\nu}(x)$ is defined as~\cite{vel97}
\begin{equation}
\tau_{\nu}(x) = 1 - \frac {\pi x^{1-\nu}}{2} [H_{\nu}(x)-N_{\nu}(x)]\; 
\label{13}
\end{equation}
where  $H_{\nu}(x)$ and $N_{\nu}(x)$, 
are the Struve and the Bessel  functions 
of order $\nu$, respectively~\cite{abr}.
The charge neutrality, Eq.~(\ref{8a}), 
together with  Eqs.~(\ref{9}) and (\ref{12}),
determines the effective charge $Q$,
\begin{equation}
Q = - \frac{  q }{\kappa_{s}a\tau_{1}(\kappa_s a) }\;.
\label{14}
\end{equation}
In the limit of $a\rightarrow 0$, the renormalized charge reduces to
the bare charge $q$.

To obtain the electrostatic free energy, we require the 
potential inside the excluded volume region,
$\phi_<(\varrho)$.  Unfortunately the procedure presented above 
leaves
this undetermined.  The only statement that we can make 
is that $\phi_<(\varrho)$ must be of the form
\begin{equation}
\phi_{<}( \varrho )= \frac{q}{D\varrho}+C(\varrho)\; .
\label{15}
\end{equation}
For a three dimensional unconstrained electrolyte, $C$ is a constant.  
This, however, is not
the case in the present geometry and $C(\varrho)$ is a function
of position. In particular the value $C(0)$ is the potential
that the central ion feels due to the presence of other
particles. In their earlier study, Velasquez and Blum~\cite{vel97} approximated
$C(\varrho)$ by a constant which they then determined by requiring 
continuity of electrostatic potential across the exclusion boundary
$\phi_{<}(a)=\phi_{>}(a)$.  This, however, is a very rough 
approximation since there is nothing to prevent $C(\varrho)$ from
being a very strongly varying function of position.  To avoid
this difficulty we shall use an alternate method of obtaining
the value of $C(0)$.  To this end we note that the potential
at the center of a circularly symmetric charge distribution is  
\begin{equation}
\label{16}
C(0;\rho)= \frac{2 \pi}{D} \int_a^{\infty} \sigma_s(\varrho) d\varrho \;\;, 
\end{equation} 
where in order to emphasize that the potential depends
on the density of ions, we have explicitly included $\rho$ in
its definition.
Using Eq.~(\ref{9}) with the potential given by Eq.~(\ref{12})
we find 
\begin{equation}
\label{16a}
C(0;\rho)= \frac{q}{Da\tau_1(k_s a)}\int_{\kappa_s a}^{\infty}\frac{\tau_0(z)}{z}dz \;\;.
\end{equation} 
 The excess chemical potential can now be calculated
straight forwardly by appealing to 
the G\"untelberg charging process~\cite{gun26}.  
We find
$\mu_{\pm}^{ex}=qC(0;\rho)/2$. The chemical potential
of the positive and negative ions is
\begin{equation}
\label{18}
\beta \mu_{\pm}=\frac{\partial f}{\partial \rho_{\pm}}=
\ln(\frac{\rho \Lambda^2}{2})+\frac{\beta qC(0;\rho)}{2}\; .
\end{equation} 
The critical point is determined by the conditions, 
$\partial \mu_{\pm}/ \partial \rho=0 $ and 
$\partial^2 \mu_{\pm}/ \partial \rho^2=0$,
which reduce to
\begin{eqnarray}
\label{20}
&2T^{*}\tau_{1}^{2}+I\tau_{1}-I\tau_{0}-\tau_{1}\tau_{0}=0 &\nonumber \\
&&\nonumber\\
&I\tau_{1}^{2}-4I\tau_{0}\tau_{1}-3\tau_{0}\tau_{1}^{2}+I\tau_{0}^{2}+2\tau_{0}^{2}\tau_{1}+x^{2}I\tau_{1}^{2}+I\tau_{1}+x^{2}\tau_{1}^{3}+\tau_{1}^{2}=0&
\end{eqnarray}
with
\begin{eqnarray}
\label{20b}
I(x)=\int_{x}^{\infty}\frac{\tau_0(z)}{z}dz \;\;,
\end{eqnarray}
where $x=\kappa_s a$.  Solving Eqs.~(\ref{20}), the critical point is located 
at $T_c^{*}=0.0517386$ and $\rho_{c+}^{*}=\rho_{c-}^{*}=0.00121988$.

\section{Debye-H\"uckel-Bjerrum approximation.}

Clearly the low temperature at which the critical point is located
should make us worry about the approximations which have been adopted.
Certainly at such extreme conditions the linearization of 
the Boltzmann factor, Eq.~(\ref{9}), is no longer valid.  Fortunately, 
all is not
lost. Evidently, linearization of the Boltzmann factor in 
Eq.~(\ref{8}) diminishes the weight of configurations
in which the oppositely charged ions are in a close proximity.
It is possible, therefore, to correct for the omitted non-linearities
by explicitly allowing for the formation
of "clusters".  These clusters are assumed to be in a chemical equilibrium
with the free unassociated ions, and their density is determined
by the law of mass action~\cite{bje26,ebe68}. The most basic such cluster is a 
dipole formed by
a $(+-)$ pair. Within the simplest version of this theory
the dipoles are treated as non-interacting ideal specie.
For $3d$ electrolyte this Debye-H\"uckel-Bjerrum approximation (DHBj)
\cite{fis93,lev96}
has proven extremely successful, predicting
the location of the critical point in close agreement with
the Monte Carlo simulations~\cite{val91}.

The total density of ions $\rho$ can then
be subdivided into that of free unassociated monopoles 
$\rho_1=\rho_++\rho_-$, and of dipolar pairs $\rho_2$,
with $\rho=\rho_1+2 \rho_2$. In the spirit of DHBj theory
we shall first treat the dipoles as ideal non-interacting entities
whose concentration is determined by the law of mass-action,
$\mu_++\mu_-=\mu_2$. The Helmholtz free energy density is then given by 
\begin{equation}
\label{21}
\beta f_{DHBj}=f_{ent}^{(2)}(\rho_1,\rho_2,T)+f_{el}(\rho_1,T)
\end{equation}
where the entropic contribution  is 
\begin{equation}
\beta f_{ent}^{(2)}(\rho_1,\rho_{2},T)=
 \left[ \rho_{2} \ln \left( \frac {\rho_{2}\Lambda^{4}}{\xi_{2} }\right) - 
\rho_{2} \right]+ \left[ \rho_{1} \ln \left( \frac {\rho_{1}\Lambda^{2}}{2} \right) -\rho_{1}\right] \;.
\label{22}
\end{equation}
\noindent The $\xi_{2}$ is the internal partition function
of a dipolar pair, 
\begin{equation}
\xi_{2}(T;R) =  2\pi \int_{a}^{R} {\bf{\mbox{\large{\em e}}}}^{- 
\beta U_2(\varrho)} \varrho d\varrho\;\;,
\label{23}
\end{equation}
\noindent where  $\beta U_2=a/T^*\varrho$ is the electrostatic 
potential between the associated ions.
In order to evaluate $\xi_{2}(T;R)$ we must define the distance
$R$ at which two ions can be considered to be associated.
Following Bjerrum~\cite{bje26} we choose the value of $R_{Bj}$ at which
$\xi_{2}(T;R)$, as a function of $R$, has an inflection point,
$R_{Bj}=a/T^*$~\cite{lev96}. With this, the integral in Eq.~(\ref{23})
can be evaluated to yield 
\begin{equation}
\xi_{2}=\frac{\pi a^{2}{{\mbox{\large{\em e}}}}^{b}}{b} 
\left[ b^{3} {{\mbox{\large{\em e}}}}^{-b} \left( Ei(b)-Ei(1)+2 
{{\mbox{\large{\em e}}}} \right) -b(1+b)         \right]\;\;,
\label{24}
\end{equation}
\noindent where $b=1/T^{*}$ and $Ei(x)$ is the exponential integral 
function. 

Since within the DHBj approximation the dipoles are treated as
ideal, the electrostatic free energy only depends on the
density of free monopoles, $\rho_1$.  Substituting the free energy
into the law of mass-action, the density of dipoles is 
given by
\begin{equation}
\rho_{2}^{*}=\xi_2 \rho_+ \rho_- e^{\beta(\mu_+^{ex}+\mu_-^{ex})}\;\;.
\label{26}
\end{equation}
The charge neutrality requires that  $\rho_+=\rho_-=\rho_1/2$.
The expression for excess chemical potential was calculated
in Section 2, $\mu^{ex}_{\pm}=\beta qC(0;\rho_1)/2$,
and the inverse Gouy-Chapman length is now 
$\kappa_{s}=2 \pi \rho^*_1/ T^* a$.

Within the DHBj approximation
the dipoles are treated as ideal specie, therefore, 
they cannot influence the location of the critical point.
Thus, the critical temperature must still be $T_c^{*}=0.0517386$
while the critical density of monopoles must remain 
$\rho_{1c}^{*}=0.00243975$. Substituting these values
into Eq.~(\ref{26}), we find that the density of dipoles
is $\rho_{2c}^{*} = 1.08847$, which is extremely high.
If there are so many dipoles is it also not possible that
there will be higher order clusters as well?

\section{Linear Ionic Chains}

Unfortunately as soon as we get to clusters of three ions
the calculations get extremely difficult.  The basic problem
is the internal partition  function of the higher order clusters, 
which can no longer be evaluated exactly. Furthermore, while it is
evident that for a cluster of three ions the low temperature
configurations are chain-like, $(+-+)$ or $(-+-)$, 
this is far from obvious for
a neutral cluster of two positive and two negative ions.  
The entropy favors a chain-like configuration $(+-+-)$, while
the energy favors a compact square cluster. Which will gain in the
critical region is hard to say.  We note, however, that
exactly the same situation was encountered for 
dipolar hard spheres~\cite{lev99,jac55}. 
In that case, in the vicinity of the critical
point, the chain like configurations dominated.  Since
it is much easier to study the linear clusters they, therefore,
will provide a starting point for our analysis.  

We begin by supposing that
at low temperatures our system will be composed of monopoles of
density $\rho_1$ and chains of $n$ monomers with densities 
$\rho_n$.  Once again
in the spirit of Bjerrum we shall first treat  the clusters
as ideal non-interacting species.  The particle conservation
requires that
\begin{equation}
\rho =\sum_{n=1}^{\infty} n \rho_{n}\;.
\label{34}
\end{equation}

Consider an alternating chain composed of 
$t$ positive and $s=n-t$ negative ions. The partition
function for such a cluster is,
\begin{equation}
\xi_{n}(T)=\frac{1}{s!t!}\frac{1}{A}\int_{\Omega_{n}}
\prod_{i=1}^{n}d^{2}{\bf r}_{i} {\bf{\mbox{\large{\em e}}}}^{-\beta U_{n}} \;.
\label{31}
\end{equation}
\noindent where $\Omega_n$ is the configurational volume  
and  $U_{n}$ is the total energy of interaction 
between the ions forming a chain, 
\begin{equation}
U_{n}({\bf r}_{1}, {\bf r}_{2}, ..., {\bf r}_{n}) =  \sum_{i < j}^{n} \varphi_{ij}(r_{ij})\; .
\label{32}
\end{equation}
At zero temperature chains are rigid and 
the particles are in contact with one another. 
The displacement vector  between two
ions $i$ and $j$ is $\vec{r}^0_{i,j}\equiv a|i-j| \hat x$,
where  $\hat x$ is the unit vector along the chain. The electrostatic
energy of interaction between the ions of the chain can be
evaluated exactly yielding
\begin{equation}
U_{n}^0 = \frac{q^{2}}{D a} S(n)\; ,
\label{36}
\end{equation}
where
\begin{equation}
S(n)=\sum_{k=1}^{n-1} \frac{(-1)^{k} (n-k)}{k}\;.
\label{37} 
\end{equation}
For nonzero but small  temperatures, deviations from $\vec{r}^0_{i,j}$
occur. These  fluctuations  can be taken 
into account by making a Taylor expansion of 
$U_n$ around the ground state up to quadrupolar order
\begin{equation}
U_n=U_n^0-\frac{q^2}{2D}\sum_{i\neq j}^n(-1)^{i+j}\{ \frac{(\vec{r}_{ij}-\vec{r}^0_{ij})_x}{(\vec{r}^0_{ij})^2}
+\frac{(\vec{r}_{ij}-\vec{r}^0_{ij})_y^2}{2|\vec{r}^0_{ij}|^3}-\frac{(\vec{r}_{ij}-\vec{r}^0_{ij})_x^2}{|\vec{r}^0_{ij}|^3}\}\; .
\label{38}
\end{equation} 
Here the subscripts $x$ and $y$ represent the components along the 
chain's direction and perpendicular
to it, respectively. 
Since we are assuming small deviations from the ground state, for each
ion $i$ we shall consider only  fluctuation contributions to $\xi_n$ 
arising from the interactions between the nearest 
and the  second nearest neighbors. Choosing as the
unit bases  $\hat x$ and  $\hat y$, vectors parallel and
perpendicular to the direction of the chain in the plane $z=0$, the  
nearest and the second nearest displacement
vectors can be written as
\begin{equation}
\label{39}
\vec{R}_1^{(i)}  = 
a(1+\lambda_i)\left(\begin{array}{r@{\quad\quad}l}\cos\phi_i\\[2ex]
\sin\phi_i\end{array}\right) 
\end{equation}
and
\begin{equation}
\vec{R}_2^{(i)}  = 
a\left[\begin{array}{r@{\quad\quad}l}(1+\lambda_i)\cos\phi_i+
(1+\lambda_{i+1})\cos\phi_{i+1}\\[2ex](1+\lambda_i)\sin\phi_i+(1+\lambda_{i+1})\sin\phi_{i+1}\end{array}\right] \;\;,
\label{40}
\end{equation} 
\noindent respectively. Here $\lambda$ is the radial and $\phi$ is 
the angular deviation from the relative positions in the ground state.

Substituting Eqs.~(\ref{39}) and (\ref{40}) into Eq.~(\ref{38}),
for small fluctuations the electrostatic energy becomes
\begin{equation}
\frac{U_{n}}{k_{B}T}=\frac{1}{T^{*}} \left \{ S(n) + \sum_{k=1}^{n-1} \lambda_{k} - \frac{1}{4} \sum_{k=1}^{n-2} \left[ \lambda_{k} + \lambda_{k+1} -\frac{1}{4}(\phi_{k}-\phi_{k+1})^{2} \right] \right \}\; .
\label{41}
\end{equation} 
In the low temperature limit, configurational integral,  Eq.~(\ref{31}), 
can  be performed explicitly yielding,
\begin{equation}
\xi_{n}(T^{*}) =\left(\frac{a^{2n-2}}{9} \right) 2^{2n}\pi^{\frac{n}{2}} 
{T^{*}}^{\frac{3n}{2}-2} {\bf{\mbox{\large{\em e}}}}^{-S(n)/T^{*}} \;\;\; 
for\;\; n \geq 3\;.
\label{42} 
\end{equation} 
\noindent The  prefactor is the result of  
thermal fluctuations while the exponential is
due to the ground state energy.

The condition of chemical equilibrium between the monopoles and 
the  n-chains is expressed 
through the law of mass action
$\mu_{n}=t\mu_+ + s\mu_-$. The chemical potential 
for  monopoles
is given by Eq.~(\ref{18}) with $\rho \rightarrow \rho_1$, 
while for n-chains the chemical potential is,
\begin{equation}
\mu_{n}=k_{B}T \ln  \left[ \frac{\rho_{n} \Lambda^{2n}}{\xi_{n}(T)} \right]
\label{30}
\end{equation}
The law of mass-action reduces to
\begin{equation}
\rho_{n}=\left( \frac{\rho_{1}}{2}\right)^{n} \xi_{n}(T^{*}) 
{\bf{\mbox{\large{\em e}}}}^{n \beta \mu_{\pm}^{ex}}
\label{33}
\end{equation} 

At the level of approximation that we have adopted,
the chains are treated as non-interacting ideal species.  
This means that just like in the case of dipoles in  Section 3, they
cannot affect location of the critical point.  Therefore 
$T_c^{*}=0.0517386 $ and  $\rho_{1c}^{*}=0.00243975$.
Substituting these values into Eq.~(\ref{33}) we find that the
sum in Eq.~(\ref{34}) diverges.  In fact according to the 
Cauchy criterion, the sum in Eq.~(\ref{34}) will converge absolutely
if and only if,
\begin{equation}
\Delta \equiv \lim_{n \rightarrow \infty} {\rho_{n}^*}^{1/n}<1\; .
\label{35}
\end{equation}
Using Eq.~(\ref{33}),
\begin{equation}
\Delta = 2\rho_1^* \sqrt{\pi T^{*3}}  
\exp\left[\frac{\alpha + I(x)/\tau_{1}(x)}{2 T^*}\right]
\label{43} 
\end{equation}
where $\alpha=2\ln2$ and $x=\kappa_s(\rho_1) a$.
Inserting the critical parameters into the expression above, 
we find that at criticality
$\Delta_c = 5.82706$, and the Cauchy criterion is strongly violated. 
Therefore, the critical density of monopoles lies outside 
the radius of convergence  of Eq.~(\ref{34}). This means that 
for any finite density $\rho$, the density of monopoles 
never reaches the threshold  necessary for the phase separation to occur,
$\rho_1<\rho_{1c}$.

\section{Results and discussion}

We have presented an argument which suggests that a confined
$2d$ electrolyte should not phase separate.  Instead, just as
for the case of dipolar hard spheres, as the temperature is
lowered, the ions will associate forming polymer-like chains
of alternating positive and negative monomers.  Clearly our
argument is based on a number of assumptions.  First, 
we have supposed
that in the critical region the linear chains
predominate over the compact clusters.  This is far
from obvious.  If the compact clusters have lower 
free energy than the chains,
they can provide nuclei for the condensation and
the gas-liquid phase separation.  A second assumption
implicit in our calculations is that the chains interact
only weakly.  This is somewhat easier to justify.  Consider 
an infinitely long rigid line of alternating charges $(...+-+-+-...)$.
Suppose that the center to center distance between the nearest
neighbors is $a$.  It is then possible to show that the potential
produced by such a line of charge decays exponentially,
$\psi(r) \sim \exp(-\pi r/a)$, where $r$ is the distance perpendicular
to the chain. Thus, the interactions between long polymer-like 
clusters should, indeed, be quite weak. However, the shorter
chains can still interact sufficiently strongly to drive  phase
transition~\cite{she95}. Finally, even
if the formation of chains prevents the liquid-gas phase separation,
it does not forestall other kinds of phase 
transitions from taking place~\cite{cam00}. At sufficiently high densities,
the two component plasma will crystallize exhibiting a 
pseudo long range order.  At low temperatures
and densities, where the polymer-like chains predominate, the Y-like
defect formation~\cite{tlu00} can lead to a coexistence between
two phases, one with  high and the other with 
low concentration of defects~\cite{pin00}.  

After this work was completed and submitted for publication,
the referee drew our attention to a recently published  
simulation by Weis, Levesque, and Caillol (WLC)~\cite{wei98} of a
$2d$  ionic fluid.  Indeed, these authors found coexistence between
high and low density phases.  
WLC estimated the critical
temperature to be $T_c \approx 0.04$, which should be compared with
our Debye-H\"uckel prediction of $T_c=0.052$.  Furthermore,
in the low density phase, WLC found predominance of chain and
ring-like clusters.  The high density phase resembles a
percolating gel-like cluster~\cite{wei98}.  Although there is a phase
coexistence, it is difficult to associate it with a traditional
liquid-vapor transition.  Instead the coexistence resembles more
a sol-gel transition in polymer systems.  The task for
theorists must now be to quantitatively understand the phase
transition found by WLC.  We hope that the current paper
will provide a first step in this direction.

\vspace*{1.25cm}
\noindent{\Large\bf Acknowledgments}

\vspace*{0.5cm} This work was supported by the 
Brazilian science agencies CNPq, FINEP, and Fapergs.
J.E.F.M. acknowledge the postdoctoral fellowship from the
CNPq and CLAF during the development 
of this work, and partial financial support from CONACyT.
M. C. B. is grateful for the hospitality
at the Center of Polymer Studies at Boston University,
where part of this work was done.


\newpage
\begin{thebibliography}{99}

\bibitem{sin90}  R.R. Singh and K.S. Pitzer,
J. Chem. Phys.  {\bf 92}, 6775, (1990)

\bibitem{wei92}  H. Weing\"artner et. al. 
J. Chem. Phys.  {\bf 96}, 848, (1992);
T. Narayanan and K.S. Pitzer, Phys. Rev. Lett {\bf 73}, 3002, (1994);
W. Schr\"oer et.al. J. Phys: Condens, Matter, {\bf 8}, 9329, (1996).

\bibitem{fis93} M. E. Fisher and Y. Levin, Phys. Rev. Lett., 
{\bf 71}, (1996) 3826.

\bibitem{lev96} Y. Levin and M. E. Fisher, Physica A {\bf 225}, (1996) 164.

\bibitem{fis96}  M.E. Fisher and B.P. Lee,
Phys. Rev. Lett.  {\bf 77}, 3561, (1996).


\bibitem{val98}  J. Valleau and G. Torrie, 
J. Chem. Phys.  {\bf 108}, 5169, (1998).

\bibitem{wei93} J.J. Weis and D. Levesque, Phys. Rev. Lett. {\bf 71},
2729, (1993); J.-M. Caillol, J. Chem. Phys, {\bf 98},
9835, (1993); M.E. van Leeuwen and B. Smit, Phys. Rev. Lett. {\bf 71},
3991, (1993); M.J. Stevens and G.S. Grest, Phys. Rev. Lett. {\bf 72},
3686, (1994)

\bibitem{sea96} R. P. Sear, Phys. Rev. Lett. {\bf 76},
2310, (1996).

\bibitem{lev99} Y. Levin, Phys. Rev. Lett. {\bf 83},
1159, (1999).

\bibitem{vel97} E. S. Velazquez and L. Blum, Physica A, {\bf 244}, (1997) 453. 

\bibitem{sal96} R.A. Salkar et.al.
J. Chem. Soc. Chem. Comm., 1223, (1996).

\bibitem{val91} J.P. Valleau, J. Chem. Phys. {\bf 95},
584, (1991); A.Z. Panagiotopoulos, Fluid Phase Equib. {\bf 76},
92, (1993); J.M. Cailol, J. Chem. Phys, {\bf 100},
2161, (1994); J.M. Cailol et. al., J. Chem. Phys, {\bf 107},
1565, (1997); G. Orkoulas and A.Z. Panagiotopoulos, 
J. Chem. Phys, {\bf 110}, 1581, (1999); Q. Yan and 
J.J. de Pablo, J. Chem. Phys, {\bf 111}, 9509, (1999) 

\bibitem{lev299} Y. Levin,  Physica A {\bf 265}, (1999) 432.

\bibitem{deb23} P.W. Debye and E. H\"uckel, Phys. Z. {\bf 24},
185, (1923);  D. A. McQuarrie, 
{\em Statistical Mechanics}, ( Harper \& Row,
 New York, 1971), Chap. 15.
 

\bibitem{abr} M. Abramowitz, I. A. Stegun (Eds.), 
Handbook of Mathematical Functions,  Dover, New York, NY, 1968.

\bibitem {gun26} E. G\"untelberg, Z. Phys. Chem. {\bf 123}, 199 (1926).

\bibitem{bje26} N. Bjerrum, Kgl. Dan. Vidensk. Selsk. Mat.-fys.
Medd. {\bf 7}, 1, (1926).

\bibitem{ebe68} W. Ebeling, Z. Phys. Chem.(Leipzig) {\bf 238},
400, (1968)


\bibitem{jac55} I.S. Jacobs and C. P. Bean, Phys. Rev.  {\bf 100},
1060, (1955).

\bibitem{she95} J.C. Shelley and G.N. Patey,
J. Chem. Phys.  {\bf 103}, 8299, (1995).

\bibitem{cam00}  P. J. Camp, J.C. Shelley, and G.N. Patey,
Phys. Rev. Lett.  {\bf 84}, 115, (2000).


\bibitem{tlu00}  T. Tlusty and S.A. Safran,  Science {\bf 290},
1328, (2000).

\bibitem{pin00}  P. Pincus, Science {\bf 290},
1307, (2000).

\bibitem{wei98} J.J. Weis, D. Levesque and J.M. 
Caillol, J. Chem. Phys. {\bf 109}, 7486, (1998)

\end{thebibliography}
\end{document}